\documentclass[figures,preprint]{epl}   
\usepackage{amsfonts}
\usepackage{amssymb}
\usepackage{amsbsy}
\usepackage{graphics}

\input epsf.sty

\newcommand{\BEQ}{\begin{equation}}     
\newcommand{\BEA}{\begin{eqnarray}}
\newcommand{\EEQ}{\end{equation}}       
\newcommand{\EEA}{\end{eqnarray}}
\newcommand{\eps}{\varepsilon}          
\newcommand{\D}{{\rm d}}                

\renewcommand{\vec}[1]{\boldsymbol{#1}} 

\title{Ageing in disordered magnets and local scale-invariance}
\shorttitle{Disordered magnets \& local scale-invariance~}

\author{Malte Henkel\inst{1,2} and Michel Pleimling\inst{3,4}}

\institute{
\inst{1}Laboratoire de Physique des Mat\'eriaux (CNRS UMR 7556), 
Universit\'e Henri Poincar\'e \\ Nancy I, B.P. 239,
F -- 54506 Vand{\oe}uvre l\`es Nancy Cedex, France\\
\inst{2}Dipartimento di Fisica/INFN - Sezione di Firenze, Universit\`a di Firenze,\\ 
I - 50019 Sesto Fiorentino, Italy\\
\inst{3} Institut f\"ur Theoretische Physik I, 
Universit\"at Erlangen-N\"urnberg,\\
D -- 91058 Erlangen, Germany\\
\inst{4} Department of Physics, Virginia Polytechnic Institute and State University,\\
Blacksburg, VA 24061-0435, USA}
\pacs{05.70.Ln}{Nonequilibrium and irreversible thermodynamics}
\pacs{64.60.Ht}{Dynamic critical phenomena}
\pacs{75.10.Nr}{Spin-glass and other random models}

\begin{document}
\maketitle
\begin{abstract}
The ageing of the bond-disordered two-dimensional Ising model quenched to below
its critical point is studied through the two-time autocorrelator and 
thermoremanent magnetization ({\sc trm}). The corresponding ageing 
exponents are determined. The form of the scaling function of the {\sc trm} 
is well described by the theory of local scale-invariance. 
\end{abstract}

A spin system quenched to below or exactly at its critical temperature 
$T_c$ undergoes ageing, that is time-translation invariance is broken
and observables such as correlation or response functions display dynamical
scaling. While systematic studies of ageing appear to have been carried out
first in studies of the mechanical response of glassy systems \cite{Stru78}, it
has been realized relatively recently that a similar phenomenology also
applies for simple ferromagnets without disorder, see 
\cite{Bray94,Godr02,Cala05,Henk06} for reviews. In these simpler systems,
for both $T=T_c$ and $T<T_c$ there is a single time-dependent 
length-scale growing with time as $L(t)\sim t^{1/z}$ where $z$ is the
dynamical exponent. $L(t)$  is 
identified with the typical linear size of correlated or ordered clusters,
respectively. It is common to study ageing through the two-time
autocorrelation function 
$C(t,s)=\langle \phi(t,\vec{r}) \phi(s,\vec{r}) \rangle$ or the 
two-time linear autoresponse function 
$R(t,s)={\delta \langle \phi(t,\vec{r})\rangle}/{\delta h(s,\vec{r})}|_{h=0}$. 
Here $\phi(t,\vec{r})$ denotes the space-time-dependent order-parameter, 
$h(s,\vec{r})$ is the space-time-dependent conjugate magnetic field, 
$t$ is referred to as {\em observation 
time} and $s$ as {\em waiting time}. For ageing systems, averages depend on 
both $t$ and $s$ and not merely on the difference $\tau=t-s$. 
In the ageing regime $t,s\gg t_{\rm micro}$ and 
$t-s\gg t_{\rm micro}$, where $t_{\rm micro}$ is some microscopic time scale, 
dynamical scaling holds ({\em at} equilibrium $a=b$)
\BEQ \label{gl:SkalCR}
C(t,s) = s^{-b} f_C(t/s) \;\; , \;\; R(t,s) = s^{-1-a} f_R(t/s)
\EEQ
such that the scaling functions $f_{C,R}(y)$ satisfy the following 
asymptotic behaviour
\BEQ \label{gl:lambdaCR}
f_C(y) \sim y^{-\lambda_C/z} \;\; , \;\; 
f_R(y) \sim y^{-\lambda_R/z}
\EEQ
as $y\to\infty$ and where $\lambda_C$ and $\lambda_R$, respectively, are known
as the autocorrelation \cite{Fish88,Huse89} and autoresponse exponents
\cite{Pico02}. For non-conserved dynamics (model A), 
these exponents are independent of the equilibrium exponents
and of $z$ \cite{Jans89}. In simple magnets without disorder and 
with short-ranged initial conditions
one finds $\lambda_C=\lambda_R$, see \cite{Bray94,Pico04} and this has been 
reconfirmed in a recent second-order perturbative analysis
of the time-dependent Ginzburg-Landau equation \cite{Maze03}. On the other hand,
for either long-ranged initial correlations in ageing ferromagnets \cite{Pico02} 
or else for disordered systems such as the random-phase sine-Gordon model \cite{Sche03},
$\lambda_C$ and $\lambda_R$ are known to be different from each other. 
Finally, if $T<T_c$, then for simple magnets with short-ranged equilibrium 
correlation functions (such as the Ising model in $d>1$ dimensions) one has 
$b=0$ and standard scaling arguments show that $a=1/z$. On the other hand, if
$T=T_c$, then $a=b=(d-2+\eta)/z$ where $\eta$ is a standard equilibrium critical
exponent.

Furthermore, according to the theory of local scale-invariance (LSI) \cite{Henk02} the
response functions of ageing systems with an algebraic growth law $L(t)\sim t^{1/z}$
are expected to transform covariantly under local space-time transformations 
$t\mapsto (1+\eps)^z t$, $\vec{r} \mapsto (1+\eps)\vec{r}$ 
with an infinitesimal $\eps=\eps(t,\vec{r})$. Restricting the admitted
transformations such that time-translations are excluded (as is natural to do 
in the context of ageing) leads to the
following prediction of the scaling function $f_R(y)$, up to normalization 
\cite{Henk02,Pico04,Henk05,Henk06a}
\BEQ \label{gl:fR}
f_R(y) =  y^{1+a'-\lambda_R/z} \left( y-1 \right)^{-1-a'}
\EEQ
where $a'$ is a new exponent. 
Over the past few years, evidence confirming the prediction (\ref{gl:fR})
has been accumulating. First, consider phase-ordering kinetics with a non-conserved
order-parameter (where $z=2$ \cite{Rute95b}). Then the prediction (\ref{gl:fR})
with $a=a'$ was confirmed in the exactly solvable random walk and the spherical 
model \cite{Godr00b} and numerically in the $2D/3D$ Ising \cite{Henk03b} and XY 
\cite{Abri04} models 
and the $2D$ $q$-states Potts model with $q=2,3,8$ \cite{Lore06}. Second, consider
non-equilibrium critical dynamics with a non-conserved order-parameter, 
where in general $z\ne 2$. Again, in many 
spin systems eq.~(\ref{gl:fR}) is either exactly reproduced in solvable models or 
describes very well the numerical data in in a large variety of systems, in general
with $a$ and $a'$ being distinct, 
including ferromagnets such as 
the Glauber-Ising model in $d=1,2,3$ dimensions, the OJK approximation, 
the spherical model and glassy systems such as the Frederikson-Anderson model and
the critical Ising spin glass, see the references quoted in \cite{Henk06a}. 
In addition, ferromagnets in restricted gometries have also been 
studied \cite{Ple04,Bau06}. Finally,
tests of eq.~(\ref{gl:fR}) for $1D$ critical systems without detailed balance 
(where (\ref{gl:SkalCR}) holds for connected correlators with $a\ne b$ in general, 
see \cite{Enss04,Rama04,Baum05,Odor06}) were
performed for the contact process (for $t/s\gtrsim 1.1$) 
\cite{Enss04,Henk06a,Hinr06}, the non-equilibrium kinetic Ising model \cite{Odor06}
and bosonic models \cite{Baum06}.  
We stress that both the theoretical prediction (\ref{gl:fR}) of LSI as well as 
almost all existing tests of it\footnote{In the
$1D$ critical contact process a full lattice was used as initial state. 
It is conceivable that this distinct initial condition might cause the departure seen 
from eq.~(\ref{gl:fR}) in the region $t/s\lesssim 1.1$ in the
contact process \cite{Hinr06,Henk06a} 
but the question is still open.}  assume an initial state with 
non-vanishing magnetization. Finally, stochastic processes such as the zero-range
process \cite{Godr06} lack the spatial structure assumed in local 
scale-invariance (for instance
Galilei-invariance for $z=2$ or a convenient generalization for $z\ne 2$) and 
do not reproduce (\ref{gl:fR}). 

In this letter, we study the ageing properties of the
$2D$ Ising model with bond disorder, but without frustrations,
which describes the ageing of diluted magnets. 
The hamiltonian is
\BEQ
{\cal H} = - \sum_{(i,j)} J_{ij} \sigma_i \sigma_j \;\; ; \;\;
\sigma_i = \pm 1
\EEQ
where the random variables $J_{ij}$ are equally distributed over the
interval $[1-\eps/2,1+\eps/2]$ with $0\leq \eps \leq 2$. We consider a square
lattice with $300 \times 300$ spins, which is large enough such that no 
finite-size effects are notable. 
It is known that this model has a second-order phase-transition and that the
value of of the critical temperature $T_c=T_c(\eps)$ does not change much 
with $\eps$, since $\langle J_{ij}\rangle =1$, 
such that $2\leq T_c(\eps)\leq 2.269$ \cite{Paul04}. 
The kinetics (non-conserved order-parameter) is described through a standard 
heat-bath algorithm. 
Throughout, we use a fully disordered initial state and average over typically
10000 different realizations of the bond disorder and of the thermal noise.
In this work, we shall limit ourselves to quenches deep into the ordered
phase $T<T_c(\eps)$. 
Recently detailed studies were carried out by Paul, Puri and Rieger 
\cite{Paul04,Rieg05} on the growth law in the low-temperature phase.
Assuming that the energy barriers for domain-wall motion depend logarithmically
on the linear domain-size $L=L(t)$, they proposed for a non-conserved order-parameter the growth law, valid for $t\to\infty$ \cite{Paul04}
\BEQ \label{gl:Lz}
L(t) \sim t^{1/z} \;\; , \;\; 
z = z(T,\eps) = 2 + {\eps}/{T}
\EEQ
and they confirmed this through extensive simulations of the
single-time correlation functions. 
It will be one of our objectives to further test this relationship. In
addition, we see that the value of the dynamical exponent can be easily changed 
so that tests of the LSI-prediction (\ref{gl:fR}) for a large range of values 
of $z$ becomes possible. This is particularly interesting since in previous 
tests of LSI in nonequilibrium critical dynamics the value of 
$z$ always remained 
quite close to $2$, the largest deviations occurring in the $1D$ 
contact process, where $z\simeq 1.6$.

\begin{figure}[t]
\centerline{\epsfxsize=3.5in\epsfbox
{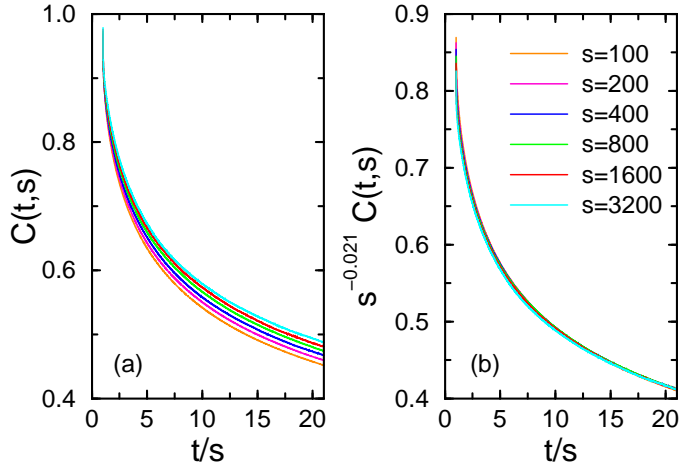}
}
\caption{Autocorrelation for $\eps=2$ and $T=1$ as a function of $t/s$ for
various waiting times $s$. For $b=0$ no data collapse is observed (a), whereas
for $b=-0.021$ the data collapse nicely for large values of $t/s$ (b). 
\label{Bild1}}
\end{figure}

We begin the discussion of our results by considering the two-times
autocorrelation function $C(t,s)$. In figure~\ref{Bild1}a we plot 
our data taken for
$\eps=2$ and $T=1$ of the autocorrelator over against $t/s$. 
In constrast to what
is found in phase-ordering for simple, non-disordered magnets, 
no data-collapse occurs here. Comparing with the anticipated long-time scaling
(\ref{gl:SkalCR}), this implies that the exponent $b$ cannot vanish, if that
scaling form is valid at all. In figure~\ref{Bild1}b we show that 
a relatively good, but not perfect scaling can be achieved with a 
non-vanishing effective value of $b$. Still,
in spite of the large waiting times used, apparently there remain important 
finite-time corrections to the anticipated scaling form (\ref{gl:SkalCR}). 
A forthcoming study \cite{Sche06a} of the site-disordered Ising model 
rather suggests 
$C(t,s) \simeq s^{-b(t/s)} f_C(t/s)$ with two continuous functions 
$f_C$ and $b$. Phenomenologically, this certainly achieves a better collapse than with a constant $b$. More detailed studies on the scaling of $C(t,s)$
will be needed. 

Assuming the standard scaling form (\ref{gl:SkalCR},\ref{gl:lambdaCR}) 
being valid for $C$, we extract the value of the autocorrelation
exponent $\lambda_C/z$, 
listed together with effective values for $b$ in  
table~\ref{Tabelle1}. 

\begin{table}
\caption[Exponenten]{Quantities describing the ageing behaviour of the disordered $2D$
Ising model.
\label{Tabelle1}}
\begin{center}
\begin{tabular}{|ll|llll|ll|l|} \hline\hline
$\eps$ & $T$ & $b$       & $a$       & $\lambda_C/z$ & $\lambda_R/z$ & $r_0$    & $r_1$     & $1/z$ \\ \hline\hline
0.5    & 1.0 & 0.005(2)  & 0.398(5)  & 0.570(5)      & 0.61(1)       & 0.021(1) & 0         & 0.400 \\
       & 0.8 & 0.001(2)  & 0.382(4)  & 0.550(5)      & 0.595(10)     & 0.020(1) & 0         & 0.381 \\
       & 0.6 & -0.008(2) & 0.353(4)  & 0.520(5)      & 0.58(1)       & 0.022(1) & 0         & 0.353 \\
       & 0.4 & -0.010(2) & 0.310(5)  & 0.460(5)      & 0.52(1)       & 0.029(2) & 0.008(1)  & 0.308 \\ \hline 
1.0    & 1.0 & -0.014(2) & 0.330(5)  & 0.490(5)      & 0.51(1)       & 0.021(1) & -0.009(1) & 0.333 \\
       & 0.8 & -0.014(2) & 0.308(4)  & 0.450(5)      & 0.49(1)       & 0.019(1) & -0.006(1) & 0.308 \\
       & 0.6 & -0.015(2) & 0.277(6)  & 0.380(5)      & 0.46(1)       & 0.020(1) & -0.007(1) & 0.273 \\
       & 0.4 & -0.013(2) & 0.22(1)   & 0.290(5)      & 0.375(10)     & 0.026(2) & -0.014(3) & 0.222 \\ \hline
2.0    & 1.0 & -0.021(1) & 0.24(2)   & 0.320(5)      & 0.33(1)       & 0.048(2) & -0.048(4) & 0.250 \\
       & 0.8 & -0.018(1) & 0.22(2)   & 0.270(5)      & 0.30(1)       & 0.093(3) & -0.042(4) & 0.222 \\
       & 0.6 & -0.012(1) & 0.17(2)   & 0.220(5)      & 0.27(1)       & 0.194(4) & -0.033(3) & 0.188 \\ \hline\hline
\end{tabular}\end{center}
\end{table}

We now turn to a discussion of the scaling of the linear response. 
While a direct
calculation of the functional derivative for $R$ would produce
extremely noisy data, it is of interest to consider an 
integrated reponse function \cite{Barr98} 
where much of the noise is smoothed out by the integration. 
Here we work with the
thermoremanent magnetization $M_{\rm TRM}(t,s)$ which is obtained after applying
a small spatially random magnetic field with amplitude $h_0=0.05$. 
{}From our experience with the scaling  of $M_{\rm TRM}(t,s)$ in the phase-ordering
of simple magnets, we expect from (\ref{gl:SkalCR},\ref{gl:lambdaCR}) the
scaling behaviour \cite{Henk02a,Henk03e}
\BEQ \label{gl:M}
M_{\rm TRM}(t,s) = h_0 \int_{0}^{s} \!\D u\: R(t,u) 
= r_0 s^{-a} f_M(t/s) + r_1 s^{-\lambda_R/z} g_M(t/s) \;\; , \;\;
g_M(y) \simeq y^{-\lambda_R/z}
\EEQ
such that the asymptotic behaviour $f_M(y)\sim y^{-\lambda_R/z}$ holds true for
$y\to\infty$. In simple magnets, the leading correction term included in
(\ref{gl:M}) often is quite sizeable and must be subtracted off before a 
reliable estimate of the scaling function $f_M(y)$ can be obtained. 

\begin{figure}[t]
\centerline{\epsfxsize=3.5in\epsfbox
{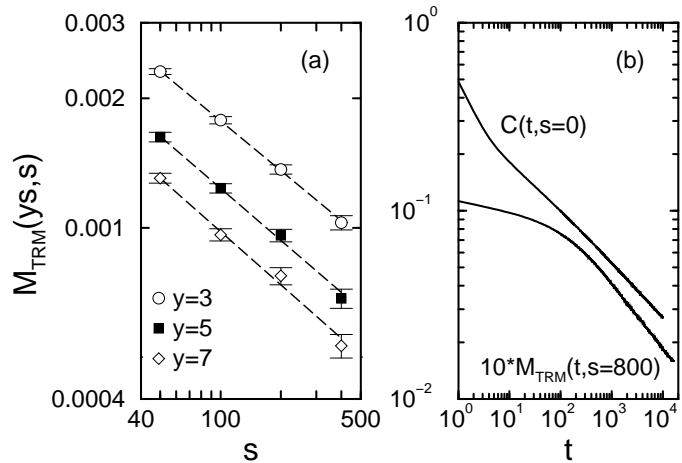}
}
\caption{(a) Thermoremanent magnetization $M_{\rm TRM}(ys, s)$ as a function
of $s$ for $\eps=0.5$ and $T=1.0$ and different values of $y$. The slopes of the
fitted straight lines yield the average value $0.398(5)$ for the exponent $a$.
(b) Time-dependent behaviour of the autocorrelator $C(t,s=0)$ and of the 
thermoremanent magnetization $M_{\rm TRM}(t, s=800)$ for $\eps=1$ and $T=0.4$.
The slopes of the straight parts are different for large values of $t$, indicating 
that the exponents $\lambda_R/z$  and $\lambda_C/z$ are different. For illustrative reasons 
$M_{TRM}$ has been multiplied by 10.
\label{Bild2}}
\end{figure}

By analogy with the simple Ising model, we determined the non-universal constants
$r_0$ and $r_1$ by fixing $y=t/s$. 
Then a fit was made from the plot of $M_{\rm TRM}(ys, s)$ over against $s$.
The results collected in table~\ref{Tabelle1} have been obtained by averaging
over at least three different values of $y$. 
As a first example, we show in figure~\ref{Bild2}a
the thermoremanent magnetization $M_{\rm TRM}(ys, s)$ for $\eps=0.5$ and $T=1$. 
It turns out that in this case
already the raw data scale very well which means that for the chosen values of the
parameters, the finite-time corrections to the leading scaling $M_{\rm TRM}(t,s)
\sim s^{-a} f_M(t/s)$ are negligible. We find the ageing exponent
$a=0.398(5)$. In phase-ordering kinetics, the dynamics of systems with 
short-ranged spatial correlators is determined
by the motion of the domain walls only \cite{Bray94}. Then simple dimensional
analysis yields the relation $a=1/z$. Comparing this with the values of $a$ for which
a data collapse is achieved, we find indeed a complete agreement with
\BEQ \label{gl:az}
a = a(T,\eps) = z(T,\eps)^{-1}
\EEQ
where $z=z(T,\eps)$ is given by eq.~(\ref{gl:Lz}), see the last column in table~\ref{Tabelle1}. 
We see that this agreement holds to within our numerical precision for all
the values of $T$ and $\eps$ we have considered. In this way, we confirm the earlier
conclusion of Rieger {\it et al.} \cite{Paul04,Rieg05} on the scaling of the
domain size $L(t)$. We also recall that for the simple Ising model, one has
$a=1/z=1/2$.\footnote{It was claimed that because of dangerous irrelevant variables,
the above-mentioned simple picture might not be valid in phase-ordering, 
leading to $a\ne 1/z$.
For example in the $2D$ Ising model $a=1/4$ was proposed \cite{Corberi}. 
Detailed studies of the simple $2D$ Ising model reconfirmed
$a=1/z=1/2$ \cite{Henk02a,Henk03b,Chat03,Henk05c,Lore06}, in agreement 
with the standard picture 
and {\it contra} the claim of \cite{Corberi}. The support 
of eq.~(\ref{gl:az}) in the diluted Ising model provides 
further evidence against the claim raised in \cite{Corberi}.} 

We point out that in contrast to the autocorrelation function, the integrated
response displays a completely conventional dynamical scaling according to
eqs.~(\ref{gl:SkalCR},\ref{gl:lambdaCR}). 

Having understood how to achieve dynamical scaling for the linear response, we 
next look at the asymptotic behaviour $f_M(y)\sim y^{-\lambda_R/z}$ for 
$y\to\infty$. 
We collect the results for the autoresponse exponent $\lambda_R/z$ in 
table~\ref{Tabelle1} and now compare with the values of $\lambda_C/z$ we 
have determined before. It can be clearly seen that the autocorrelation and
autoreponse exponents are different for the model at hand. We further 
illustrate 
this in figure~\ref{Bild2}b where the autocorrelation function 
$C(t,0)\sim t^{-\lambda_C/z}$ for $t$ large enough and the 
thermoremanent magnetization $M_{\rm TRM}(t,800)\sim t^{-\lambda_R/z}$ 
are compared for $\eps=1$ and $T=0.4$ in a log-log plot. Clearly the slopes
for $t$ large and hence the two exponents are different. This behaviour sets
the kinetics of the bond-disordered Ising model apart from what is known for 
the phase-ordering of all simple magnets with short-ranged initial 
correlations. 
We did not anticipate this finding but recall that the random-phase sine-Gordon
model also shows $\lambda_C\ne \lambda_R$ slightly below its 
critical temperature \cite{Sche03}. 

\begin{figure}[t]
\centerline{\epsfxsize=5.0in\epsfbox
{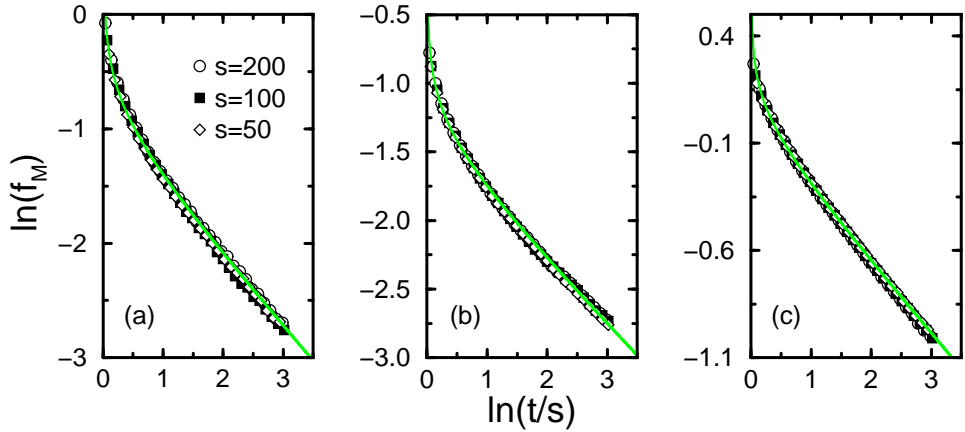}
}
\caption{Comparison of the numerically determined scaling functions
$f_M(t/s)$ with the predictions coming from local scale-invariance (full lines).
(a) $\eps=0.5$ and $T=1$, (b) $\eps=1$ and $T=0.6$, (c) $\eps=2$ and $T=1$.
In the first case corrections to scaling are negligibly small.
Error bars are smaller than the sizes of the symbols.
\label{Bild3}}
\end{figure}

After these preparations, we are ready to study in more detail the form of the
scaling function $f_M(y)$ of the integrated linear response. It is straightforward 
to integrate the prediction (\ref{gl:fR}) of local scale-invariance with 
$a=a'$. We have (${}_2F_1$ is a hypergeometric function)
\BEQ \label{gl:LSI}
f_{M}(y) = y^{-\lambda_R/z} {}_2F_1\left( 1+a, \frac{\lambda_R}{z}-a;
\frac{\lambda_R}{z}-a+1; \frac{1}{y} \right)
\EEQ
In this expression, the values of the exponents and also the normalization 
are already fixed, see table~\ref{Tabelle1}. We can now compare the data for 
the scaling function with the prediction (\ref{gl:LSI}) of LSI. 
In figure~\ref{Bild3} examples for this comparison are shown for three
choices of $\eps$ and $T$. In the first case we have already seen that 
finite-time corrections to scaling were negligibly small and we now 
see from figure~\ref{Bild3}
that the form of the scaling function is perfectly described by LSI. 
In the two other cases, the finite-time corrections are notable and actually 
quite sizeable for $\eps=2$. Subtracting them, a very good scaling 
behaviour is found. Again, the curve (\ref{gl:LSI}) matches the data 
very well. This agreement with LSI could not hold if 
(\ref{gl:Lz}) or (\ref{gl:az}) were invalid. 

It is remarkable, in view of the very large values of $z$, how well 
the theory of
local scale-invariance describes the data. From a 
field-theoretical perspective, 
it was advanced \cite{Cala05} that a symmetry principle such as local 
scale-invariance could at best hold at the mean-field level, where $z=2$. 
The present example, using a non-integrable
spin system with values of the dynamical exponent which can become very large,
rather suggests that the idea of extending dynamical scaling to a more local
form should indeed be capable of faithfully reproducing at 
least the linear responses
of physically quite distinct systems. At present, it appears that the main
requirements on the models to be studied are an algebraic 
growth law $L(t)\sim t^{1/z}$
of the linear size of the correlated clusters and a 
vanishing initial magnetization. 
The present study closes a gap between the simple, non-disordered magnets and
on the other side {\em critical} spin glasses \cite{Henk05} where 
local scale-invariance has been successfully tested before.

In summary, we have studied the ageing behaviour of a 
bond-disordered two-dimensional
Ising model quenched to temperatures far below its critical temperature
$T_c(\eps)$. While, the linear response as studied by 
the thermoremanent magnetization
was seen to be completely compatible with standard dynamical scaling, large
finite-time corrections to standard dynamical scaling were seen in the two-time 
autocorrelation function. Our data appear to be compatible with
the standard relation $a=1/z$ of the ageing exponent $a$ with 
the dynamical exponent 
$z(T,\eps)$ given by eq.~(\ref{gl:Lz}). We also encountered two surprises:
first, assuming (\ref{gl:SkalCR}), the effective ageing exponent $b$ was 
found to be non-vanishing (and negative) 
and second, the autocorrelation and autoresponse exponents are 
different, $\lambda_C\ne \lambda_R$. 
In spite of the large range of values of $z$, the form of the scaling function
of the thermoremanent magnetization is described to within the 
numerical accuracy by the theory of local scale-invariance.

\acknowledgments
MH thanks the Dipartamento di Fisica/INFN di Firenze for warm hospitality.
This work was supported by Procope, 
by CINES Montpellier (projet pmn2095), and by the Deutsche Forschungsgemeinschaft
(grant no. PL 323/2).



\end{document}